\documentclass[12pt]{iopart}

\usepackage[dvips]{graphicx,color}

\begin{document}

\newcommand{\la}{\langle}
\newcommand{\ra}{\rangle}
\newcommand{\vc}{\mathbf}
\newcommand{\bS}{{\bf{S}}}
\newcommand{\bK}{{\bf{K}}}
\newcommand{\bk}{{{\bf{k}}}}
\newcommand{\bq}{{\bf{q}}}
\newcommand{\ha}{{\hat{a}}}
\newcommand{\hb}{{\hat{b}}}
\newcommand{\hc}{{\hat{c}}}
\newcommand{\upa}{{\uparrow}}
\newcommand{\dna}{{\downarrow}}
\newcommand{\nn}{{\nonumber}}
\newcommand{\be}{\begin{equation}}
\newcommand{\ee}{\end{equation}}
\newcommand{\bea}{\begin{eqnarray}}
\newcommand{\eea}{\end{eqnarray}}
\renewcommand{\ni}{\noindent}

\title{$SU(N)$ Quantum Spin Models: A Variational Wavefunction
Study}

\author{Arun Paramekanti}
\address{Department of Physics, University of Toronto, Toronto, Ontario M5S-1A7 CANADA}
\ead{\mailto{arunp@physics.utoronto.ca}}

\author{J. B. Marston}
\address{Department of Physics, Box 1843, Brown University, Providence, RI 02912-1843 USA}
\ead{\mailto{marston@physics.brown.edu}}

\begin{abstract}
$SU(N)$ quantum spin systems may be realized in a variety of physical
systems including ultracold atoms in optical lattices.  The study of such 
models also leads to insights into possible novel quantum phases and
phase transitions of $SU(2)$ spin models.
Here we use Gutzwiller projected fermionic variational wavefunctions to explore 
the phase diagram and correlation functions of $SU(N)$ quantum spin models
in the self-conjugate representation. In one dimension,  the
ground state of the $SU(4)$ spin chain with Heisenberg bilinear and biquadratic interactions
is studied by examining instabilities of the Gutzwiller projected
free fermion ground state to various broken symmetries.
The variational phase diagram so obtained agrees well with exact results.  
The spin-spin and dimer-dimer correlation functions
of the Gutzwiller projected free fermion state with $N$ flavors of fermions
are in good agreement with exact and $1/N$ calculations for the
critical points of $SU(N)$ spin chains. 
In two dimensions, the phase diagram of the antiferromagnetic Heisenberg model
on the square lattice is obtained by 
finding instabilities of the Gutzwiller projected $\pi$-flux state. 
In the absence of biquadratic interactions the model exhibits
long range N\'eel order for $N=2$ and $4$, and spin Peierls (columnar dimer) 
order for $N > 4$.
Upon including biquadratic interactions in the $SU(4)$ model (with sign appropriate to a fermionic Hubbard model), the N\'eel
order diminishes and eventually disappears, giving way to an extended valence 
bond crystal.
In the case of the $SU(6)$ model, the dimerized ground state melts at sufficiently
large biquadratic interaction yielding to a projected $\pi$-flux spin liquid phase which
in turn undergoes a transition into an 
extended valence bond crystal at even larger biquadratic interaction.
The spin correlations of the projected $\pi$-flux state at $N=4$
are in good agreement with $1/N$ calculations. We find that the state
shows strongly enhanced dimer correlations, in qualitative agreement
with recent theoretical predictions. We also compare our 
results with a recent quantum Monte Carlo study of the 
$SU(4)$ Heisenberg model. 
\end{abstract}



\maketitle

\section{Introduction}

Quantum spin models provide a setting in which one can explore interesting
strong correlation physics that arises from quantum fluctuations. Such fluctuations can be large enough, in certain cases, to melt any
form of classical order leading to various exotic spin liquid states\cite{fazekas74,misguich03}.  A class of spin liquids of particular interest are those that support gapless excitations.  
Apart from experimental questions regarding their possible existence
\cite{coldea01,coldea03,shimizu03,kurosaki05,robert06}, 
it is interesting to inquire under what circumstances such gapless spin liquids can arise in two or more dimensions.  The Bethe ansatz solution of the spin-1/2  Heisenberg antiferromagnet chain is a well-known exact  example of such a gapless spin-liquid in one spatial dimension, but much less is known about higher dimensions.  This question was raised in the early days of high-T$_c$ theory, as Anderson's original proposal for a resonating valence bond (RVB) spin liquid had a Fermi surface of gapless spinon excitations\cite{anderson87,bza}.

One route to accessing the physics of some of these spin disordered states 
is by generalizing the usual $SU(2)$ spin models to $SU(N)$ models.  As $N$ increases from 2 to larger values, quantum spin fluctuations are enhanced, weakening any spin order.  Such models can 
accommodate several types of spontaneously broken symmetries:  Spin order, 
spin dimerization, and charge-conjugation symmetry breaking\cite{arovas}.   
We note that models of $SU(N)$ quantum spins are not purely theoretical 
exercises:  It may be possible to realize them with ultracold atoms in
optical lattices\cite{honerkamp04,hofstetter05,buchler05},
in quantum dot arrays\cite{onufriev}, or as special points in models with spin and orbital degrees of freedom\cite{mila2003}, although the effects of $SU(N)$ symmetry breaking terms
need to be carefully examined in each case.
 
Of the many possible representations of $SU(N)$, we focus on a particular self-conjugate representation  with $N/2$ fermions on each site, each with a different flavor due to the Pauli exclusion principle\cite{AM,MA}.  In this representation, the generators of the $SU(N)$ algebra may be expressed in terms of the fermion creation and annihilation operators as: 
$S^\alpha_\beta(i) \equiv f^{\dagger \alpha}_{i} f_{i \beta} - \frac{1}{2} 
\delta^\alpha_\beta$.  The constraint 
$S^\alpha_\alpha(i) = 0$ thus holds in the subspace with exactly $N/2$ fermions on each site, and consequently there are the correct number ($N^2-1$) of special unitary generators.   The representation is called ``self-conjugate'' because upon making a particle-hole transformation the same representation, namely one with $N/2$ fermions, is obtained.  All representations of $SU(2)$, regardless of the total spin, are automatically self-conjugate, but only certain representations of $SU(N>2)$ are self-conjugate.  An advantage of the self-conjugate representation is that it is easy to construct $SU(N)$ invariant Hamiltonians that retain all of the symmetries of the underlying lattice, and hence mimic $SU(2)$ models in this regard.  For more details, including the Young tableau classification, see reference \cite{arovas}.
 
In the $N \rightarrow \infty$ limit saddle point solutions of the fermionic path integral are exact and the operator constraint $S^\alpha_\alpha(i) = 0$ constraint can be replaced by the much simpler mean-field constraint $\la S^\alpha_\alpha(i) \ra =0$. 
These large-$N$ $SU(N)$ antiferromagnets cannot break global $SU(N)$ spin symmetry; some possess ground states that do not break any lattice symmetry and thus furnish mean field caricatures of a class of spin liquids. Going from $N=\infty$ down to the physical limit of $N=2$ requires the inclusion
of fluctuations about the mean field state, and the problem can be recast 
in the form of a strongly coupled gauge theory interacting with fermionic
matter fields. Progress toward calculating the properties of such a field 
theory relies on a $1/N$ expansion, and some results have been obtained
in this manner for one and two dimensional models\cite{marston8,marston9,kim-lee,rantner,vafek02,franz03,herbut03a,herbut03b,kaveh05,hermele04,tanaka05,hermele05,nogueira05,sslee05,ghaemi05} though not without controversy.  There is also good reason to be concerned about the reliability of such an expansion
in the physically important $SU(2)$ case as $N$ is of course no longer large.  DMRG calculations for $SU(N)$ quantum antiferromagnets\cite{onufriev} and Hubbard models\cite{solyom06} provide some confirmation of analytical understanding of one-dimensional (1D) chains.  In two dimensions (2D) there is a very interesting quantum Monte Carlo (QMC) study of $SU(N)$ quantum antiferromagnets by Assaad\cite{assaad05} that we discuss further in section \ref{2d}. 

A different approach to the study of $SU(N)$ spin models also begins with mean field states that satisfy the average constraint $\la S^\alpha_\alpha(i) \ra =0$, but then the operator constraint
$S^\alpha_\alpha(i) =0$ is implemented via an 
on-site Gutzwiller projection. The Gutzwiller projection operator forces the number 
of fermions to be precisely $N/2$ at each site. For $N=2$, and denoting as usual 
the two spin states by $\upa,~ \dna$, the projection operator takes the well-known
form $P_G = \prod_i (1-n_{i \upa} n_{i \dna})$.  The resulting many-body
 wavefunction thus lives in the correct Hilbert space for $SU(N)$ antiferromagnets 
and serves as a variational approximation to the spin ground state.   For $N=\infty$, a 
probabilistic central limit argument shows that
the Gutzwiller projection is unimportant and gives the same result
as the mean field state.  For any finite $N$, projection is
crucial and nontrivial; the advantage of the approach is
that the projection constraint can be handled numerically exactly with 
the variational Monte Carlo algorithm.  
Thus the generalization to $SU(N)$ provides a rationale for understanding the 
Gutzwiller procedure:  Mean-field wavefunctions obtained in the large-$N$ limit 
are then modified by projection to account for the occupancy constraint at finite-$N$.  
The approach however suffers from the criticism that it is biased and 
restricted by the choice of the variational mean 
field state (or equivalently the preprojected wavefunction) and that 
{\it local} Gutzwiller projection may not account for all of the important correlations. 

The Gutzwiller variational approach has been applied to a variety of $SU(2)$ quantum antiferromagnets; see for instance Refs. \cite{zhang88,gros89,hsu,paramekanti01,paramekanti04,sorella03,sorella06,gan05,motrunich05}.  Many of these studies support the possibility of 2D spin liquids with gapless spin excitations.  One advantage of extending the Gutzwiller variational approximation to $SU(N)$ quantum antiferromagnets is that the approximation becomes more accurate as $N$ increases.  Indeed, since the gauge theory approach and the variational approach reduce to the same (exact) mean field theory at $N=\infty$, but suffer different criticisms at finite $N$, it is interesting to compare results from both approaches
as $N$ is decreased systematically from $N=\infty$
down to $N=2$.  The existence of some exact results for $SU(N)$ quantum antiferromagnets in one and higher spatial dimensions\cite{arovas} provides valuable checks not available in the $SU(2)$ case.  So motivated, we numerically explore in this paper Gutzwiller wavefunctions for $SU(N)$ spin models in the self-conjugate representation with Heisenberg bilinear and biquadratic interactions.   
In section \ref{1d} we compare the variational approach in 1D with exact results and analytical $1/N$ calculations.  Section \ref{2d} focuses on the 2D
square lattice.  Comparison is made with $1/N$ calculations
and with Assaad's quantum Monte Carlo results.  
We summarize and discuss the 
implications of our results in section \ref{discuss}. 
For the reader interested in the  main results, phase diagram figures in the different sections provide a quick overview of our conclusions.  

\section{One Dimension}
\label{1d}

We begin with a study of $SU(N)$ spin models in 1D.  The existence of a reliable phase diagram for the $SU(4)$ spin chain\cite{arovas} provides a valuable check on the quality of the variational wavefunctions.  We first define the model and then compare the phase diagram of the $SU(4)$ chain as obtained with the variational wavefunctions to the known result.  Then we examine 
various correlation functions calculated from the Gutzwiller projected $SU(N)$
Fermi gas wavefunction 
and compare the exponents so obtained to exact analytical results for the 
critical point in the $SU(N)$ spin chain.

\subsection{Models}

For $N=2$, the usual Heisenberg model is the only nearest-neighbor Hamiltonian that is both $SU(2)$ symmetric and translationally invariant.  In 1D we may write it as:
\be
H_2 = \sum_{i} S^\alpha_\beta(i) S^\beta_\alpha(i+1),
\ee
where the trace ${\rm Tr} [S(i) S(i+1)]$ is simply
a rewriting of the vector form $H_2 = 2 \sum_i \vec{S}_i \cdot \vec{S}_{i+1}$ 
with standard spin operators $\vec{S}_i = \frac{1}{2} f^{\dagger \mu}_{i} 
\vec{\sigma}^\nu_\mu f_{\nu i}$ in terms of the matrix form of the spin operators. 
It is well-known that the model has no long range order, but rather exhibits 
algebraically decaying antiferromagnetic spin correlations up to a mutiplicative 
logarithmic factor.

For $N=4$, the most general translationally-invariant nearest-neighbor Hamiltonian can 
have an additional biquadratic spin-spin interaction term:
\be
H_4 = \cos\theta \sum_{i} S^\alpha_\beta(i) S^\beta_\alpha(i+1)
+ \frac{\sin\theta}{4} \sum_{i} [S^\alpha_\beta(i) S^\beta_\alpha(i+1)]^2,
\label{su4}
\ee
where $-\pi < \theta \leq \pi$ parametrizes the relative strength of the 
Heisenberg and biquadratic terms.  Whereas the usual bilinear Heisenberg interaction 
exchanges two fermions on adjacent sites, the biquadratic term exchanges two pairs of 
fermions.  The antiferromagnetic region of the phase diagram of $H_4$ is generically 
gapped \cite{MA,arovas,solyom05}.  The Lieb-Schultz-Mattis theorem then says that 
these gapped phases must break translational symmetry in one way or another
\cite{affleckLieb86}. Tuning $\theta$ leads to a variety of 
phases and phase transitions. For instance, positive $\theta$ frustrates dimerization, 
and when $\theta$ becomes sufficiently large, the dimerized phase is eliminated.

\subsection{Phase Diagram of the $SU(4)$ Model}

\begin{figure}
\centerline{\includegraphics[width=5.0in]{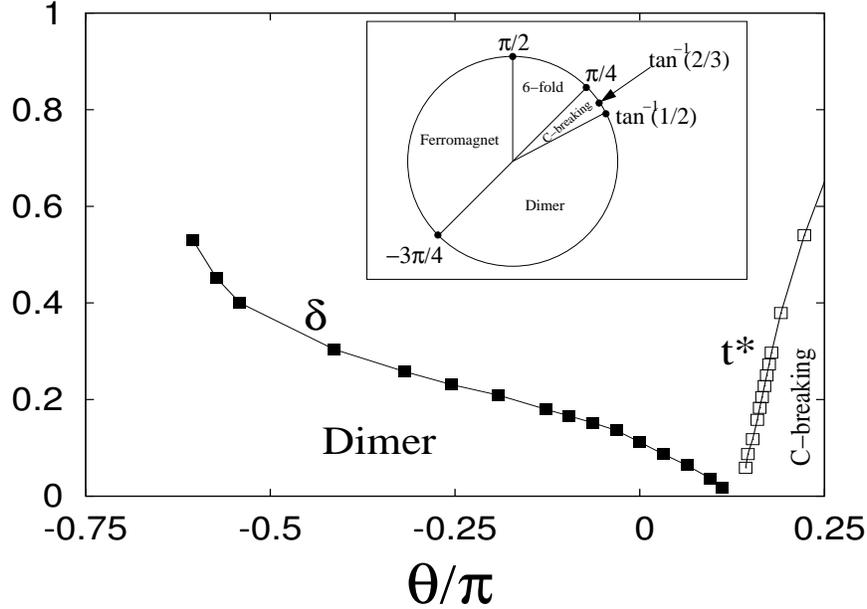}}
\caption{Variational calculation of the ground state energies of dimerized 
${\cal D}$
and charge-conjugation broken (${\cal C}$) phases of the
model Hamiltonian $H_4$ in equation \ref{su4}. Both broken symmetries apparently 
vanish around $\theta = 0.42
(3)$, hinting at a possible continuous phase transition with a 
gapless critical point that is reasonably well-described by the Gutzwiller 
projected 4-flavor Fermi gas wavefunction (see text). 
\underline{Inset:} The full phase 
diagram, including the ferromagnetic and 6-fold symmetry broken phases,
of the $SU(4)$ spin chain with biquadratic interactions.
The exact location of the ${\cal D}-{\cal C}$ transition is at 
$\theta = \tan^{-1}(1/2)
\approx 0.4636$. The point at which the ${\cal C}$ extended valence bond product state is an exact 
ground state is $\theta = \tan^{-1}(2/3)$.}
\label{fig1}
\end{figure}

The phase diagram of the $SU(4)$ model, equation \ref{su4},  was
obtained in reference \cite{arovas} and is shown in the inset to figure \ref{fig1}. 
 It displays
four phases: A fully polarized ferromagnet (${\cal FM}$), a dimerized phase (${\cal D}$), a 
phase with broken charge-conjugation symmetry (${\cal C}$), and, finally, a phase with 
broken translational symmetry and a 6-site unit cell (6-fold). 
The ${\cal D}$ to ${\cal FM}$ transition and the 6-fold 
to ${\cal FM}$ transition are both first order, while the ${\cal D}$ to
${\cal C}$ and the ${\cal C}$ to 6-fold state transitions are both
continuous.  We now describe these phases more precisely and present the 
results of calculations based upon the corresponding variational wavefunctions.

\noindent \underline{${\cal FM}$, Ferromagnet:}
The ground state of a fully polarized ferromagnet breaks the 
global $SU(4)$ spin symmetry but no lattice symmetries. A simple and exact
ground state wavefunction may then be constructed by placing any two of the four 
fermion flavors on the lattice, each site having the same two flavors. All 
other ground states can be obtained by global $SU(4)$ rotations of the 
state. Because the Pauli exclusion principle prevents any fermion hopping in the
${\cal FM}$ state, it is straightforward to show that 
$\la S^\alpha_\beta(i) S^\beta_\alpha(i+1) \ra = 
\la [S^\alpha_\beta(i) S^\beta_\alpha(i+1)]^2 \ra = 1$,
so that the exact ground state energy per site is 
\be
e^{\rm exact}_{\cal FM} = 
\cos\theta + \frac{1}{4} \sin\theta.
\ee

\noindent \underline{${\cal D}$, Dimerized:}
The dimerized phase is a spin gapped phase with a two fold
ground state and a nonzero order 
parameter $\delta_{\cal D} = \la S^\alpha_\beta(i) S^\beta_\alpha(i+1)
\ra - \la S^\alpha_\beta(i-1) S^\beta_\alpha(i) \ra$ that takes on
equal positive/negative values in the two ground states and is zero
in a state with no broken symmetries. This phase does not break the 
$SU(4)$ spin symmetry, but does
break lattice translations (the ground state is invariant only under
translation by two lattice spacings), and inversion symmetry about 
a lattice {\it site}. In order to obtain a wavefunction for this phase, we
consider a mean field fermion Hamiltonian with alternating hopping
strengths $(1+\delta)$ and $(1-\delta)$ on successive bonds:
\be
H^0_{\cal D} = - \sum_{i,\alpha=1\ldots 4} (1+\delta (-1)^i) 
\left[f^{\dagger \alpha}_i f_{i+1,\alpha} + {\rm h. c.} \right]\ .
\ee
This starting Hamiltonian has the following favorable features:
it is $SU(4)$ symmetric, has a gap to fermion excitations (and thus a gap 
to spin excitations in mean field theory), 
and breaks the same lattice symmetries as the 
dimerized phase. In addition, the ground state of the Hamiltonian
satisfies $\la S^\alpha_\alpha(i) \ra_0 =0$ since it is particle-hole
symmetric. The ground state of this mean field model is simply
a product of four Slater determinants, one for each flavor of fermion, with
the lowest half of the single particle states filled.  Gutzwiller
projecting the mean-field ground state leads to a variational ansatz for the
dimerized phase of the spin model, with $\delta$ being the variational
parameter that we optimize by minimizing $\la H_4 \ra$ to find the 
best variational ground state. We find that the dimerization
strength, defined in the mean field problem via the variational
parameter $\delta$, is nonzero at $\theta=0$ indicating that the
$SU(4)$ Heisenberg model has a dimerized ground state. With increasing
$\theta$, the optimal $\delta$ decreases and vanishes around $\theta 
\approx 0.41(3)$, in reasonably good agreement with the exact
result $\theta_c = \tan^{-1}(1/2) \approx 0.4636$. We also find that 
$\delta$ increases in magnitude for negative
$\theta$ and approaches unity at large negative $\theta$ indicating that dimerization is
nearly complete. 

\noindent \underline{${\cal C}$, Charge-conjugation symmetry broken:}
The model $H_4$ is invariant under the global particle-hole transformation 
$f_{i\alpha} \leftrightarrow f^{\dagger \alpha}_i$ and thus possesses charge-conjugation (C) invariance.
In terms of the spin operators, the transformation takes the form
$S^\alpha_\beta(i) \to - S^\beta_\alpha(i)$.  The ${\cal C}$ phase corresponds
to a phase in which this symmetry is spontaneously broken. 
This phase does not break the $SU(4)$ spin symmetry but
does break lattice symmetries as the ground states are invariant
only under translation by two sites, and they break inversion symmetry
about the {\it bond centers}.  It is characterized by an order parameter made up of a triple
product of spins:  $\langle Tr (S_i S_{i+1} S_{i+2}) \rangle$.  An extreme caricature of the state 
(analogous to the product state of nearest neighbor dimers) is an extended valence bond
solid of site-centered $SU(4)$ singlets formed from two flavors of fermions at a central site combined with a fermion from each of the two flanking sites.  This product state is an exact
ground state of model $H_4$ at the special point $\theta=\theta^*=\tan^{-1}(2/3)$.

The mean field Hamiltonian we use to
obtain the preprojected wavefunction for the C-breaking state is
\be
H^0_{{\cal C}}= 
- \sum_{i,\alpha=1\ldots 4} \left[f^{\dagger \alpha}_i f_{i+1,\alpha} 
+ {\rm h. c.} \right]
-  t^* \sum_{i,\alpha=1\ldots 4} (-1)^i 
\left[f^{\dagger \alpha}_i f_{i+2,\alpha} + {\rm h. c.} \right]\ .
\label{c-break}
\ee
Nonzero $t^*$ breaks the global particle-hole symmetry since it
connects sites belonging to the same sublattice. The alternating
sign of $t^*$ on the odd and even sublattices 
breaks inversion symmetry about bond centers of the lattice, and
generates a gap in the mean field fermion spectrum (and thus a 
gap to spin excitations in the mean field theory). Finally, the
Hamiltonian is invariant under a global particle hole transformation
followed by translation by one lattice spacing, and hence satisfies
$\la S^\alpha_\alpha(i) + S^\alpha_\alpha(i+1) \ra_0 =0$. We can further
modify the wavefunction to include a staggered chemical potential in
order to obtain $\la S^\alpha_\alpha(i) \ra_0 =0$ at each site. However since 
the mean field state is projected into the correct Hilbert space that satisfies
$S^\alpha_\alpha(i) =0$ exactly, we choose to work with the simpler mean field
Hamiltonian without this staggered chemical potential (we have checked
that including the staggered chemical potential does not affect the
results in any significant quantitative manner). 

A check on the quality
of this variational wavefunction for the ${\cal C}$ phase is provided by a 
comparison between the variational and exact ground state energy
per site at the point $\theta^*=\tan^{-1}(2/3)$ where a ${\cal C}$
product is the exact ground state while the variational ground state
exhibits nonzero $t^*$.  The variational ground
state energy at this point, $E_{\rm var}(\theta^*) = -0.6934(10)$,
is close to the exact result $E_{\rm exact}(\theta^*) = 
-\frac{5}{2\sqrt{13}} \approx -0.69337\ldots$.

Turning to general $\theta$, we find that the variational parameter 
$t^*$ is zero for $\theta < 0.42(2)$, and increases monotonically
for $\theta > 0.42(2)$. The $\theta$
at which $t^*$ first becomes nonzero thus seems to coincide (within
numerical error) with the point where the dimerization parameter 
$\delta$ vanishes ($\theta = 0.41(3)$). 

\noindent \underline{6-fold degenerate state:}
The phase diagram of the $SU(4)$ spin chain has two special points with enlarged SU(6) symmetry, and a gapped 6-fold degenerate phase is associated with one of these points\cite{arovas}.  We may understand the origin of the 6-fold degeneracy as follows:  View the 6 possible states of the self-conjugate SU(4) representation (two distinct flavors chosen out of four possible ones) on each site as the six  states of the fundamental representation of $SU(6)$.  Six such states, taken from six adjacent sites, can be combined into a $SU(6)$ singlet. The resulting spin-gapped ground
state breaks translational symmetry with a 6-site periodicity.  Unfortunately 
we have not yet found a way to express the state in terms of the Slater 
determinants of single-particle wavefunctions, 
a proper description likely needs some form of pairing to capture the above 
physics.  We therefore do not focus on this phase at present, postponing it to future 
study.

\noindent \underline{${\cal D}-{\cal FM}$ transition}
We know from the phase diagram of reference \cite{arovas} that there is a first order
${\cal D}$ to ${\cal FM}$ transition. Since the dimerization order
parameter appears to increase monotonically at larger negative
values of $\theta$, we may estimate 
the approximate 
location of the ${\cal D}-{\cal FM}$ transition by comparing the 
energy of the
${\cal FM}$ state with a {\it fully}
dimerized state ($\delta=1$) that is
just a product of nearest neighbor singlets on alternate bonds. The
energy of this variational state can be found knowing that for 
neighboring uncorrelated sites (not on the same singlet bond) 
$\la S^\alpha_\beta(i) S^\beta_\alpha(i+1) \ra = 0$
and $\la [S^\alpha_\beta(i) S^\beta_\alpha(i+1))]^2 \ra = 5/3$, while
for neighboring spins that form the singlet bond
$\la S^\alpha_\beta(i) S^\beta_\alpha(i+1) \ra = -5$
and $\la [S^\alpha_\beta(i) S^\beta_\alpha(i+1)]^2 \ra = 25$. This
leads to
\be
e^{\delta=1}_{\cal D}
= - \frac{5}{2} \cos\theta + \frac{10}{3} \sin\theta.
\ee
Comparing this energy with $e^{\rm exact}_{\cal FM}$, we find the variational
estimate of the angle at which
the dimer state becomes unstable to ferromagnetism to be 
$\theta^{\rm var}_{\cal D, 
\cal FM} = \tan^{-1}(84/74) \approx -0.73\pi$, very 
close to the exact result $\theta^{\rm exact}_{\cal D-\cal FM}
= - 3\pi/4$. The error in the variational estimate of the
transition angle is consistent with the fact that while the
energy of the ferromagnet is obtained exactly, the energy of the
dimer product state with $e^{\delta=1}_{\cal D}$ 
is only an upper bound to the true ground state energy of the dimerized 
state at the transition.

\noindent \underline{${\cal C}-{\cal D}$ transition}
The numerical coincidence of the values of $\theta$ at which $\delta$
and $t^*$ vanish indicates that the 
transition between the ${\cal D}$ and ${\cal C}$ phases could be 
continuous even within the
variational approach as in the rigorous phase diagram. We have not studied 
wavefunctions with coexisting
$t^*$ and $\delta$ broken symmetry parameters and cannot rule out the possibility
that such a variational ansatz may have lower energy in the
region close to the transition. We also cannot rule out the possibility that the variational
approach leaves a very small
window where $t^*=\delta=0$ giving rise to a gapless
phase. Assuming however that neither of these possibilities is realized,
and that the transition between the two
phases is continuous even within the variational approach, 
the projected half-filled Fermi gas state (with $\delta=t^*=0$ in 
the mean field Hamiltonian) is a good candidate for the critical point describing 
the ${\cal C}-{\cal D}$ transition. We turn next to
a study of correlation functions of this wavefunction.

\subsection{Correlation Functions for the Projected Fermi Gas at Various $N$}

As the phase diagrams of the $SU(2)$ and $SU(4)$ spin chains are known\cite{arovas}, these provide good test cases for the method.     
At the critical point in the $SU(N)$ chain, the exponents of the spin-spin and 
dimer-dimer correlation functions follow directly from equations (4.10) and (4.12) of reference \cite{arovas}, once the scaling dimension of the level 
$k = 1$ Wess-Zumino-Witten (WZW) field $g$ is known.  (The contribution of the 
currents $J_L$ and $J_R$ in equation (4.10) to the spin-spin correlation function 
is subleading, as the currents have dimension $1$, greater than that of the 
g-field.)   Reference \cite{bouwknegt99} gives the dimension of a tower of WZW operators labeled by the integer $a$:
\begin{eqnarray}
{\rm dim}(g) &=& h + {\bar h}
\nonumber \\
h = {\bar h} &=&  \frac{a (N - a)}{2N},\ a = 1, 2, \ldots, N,
\label{dimg}
\end{eqnarray}
with $a=1$ corresponding to the operator with smallest nontrivial dimension.
The conformal charge $c = N - 1$ is correct as it equals the number of 
fermions in the corresponding $SU(N)$ Hubbard model minus 1 due to the 
freezing out of charge fluctuations, leaving only $N-1$ spin excitations.   
Now the exponent of the staggered part of the spin-spin correlation function, 
as well as the dimer-dimer correlation function, is twice the 
dimension of $g$, yielding $2 - 2/N$.  This exponent
reduces to the free fermion value of $2$ in the $N \rightarrow \infty$ 
limit, and to $1$ in the usual $SU(2)$ Heisenberg chain, as it should.  

How do correlation functions behave in the projected Fermi gas wavefunction
for general $N$? For $N=2$ the Gutzwiller projected Fermi gas wavefunction at half-filling 
is the exact ground state of the Heisenberg model with $1/r^2$ interactions
\cite{haldane,shastry}. It also correctly describes correlation functions of the 
ground state of the $J1-J2$ Heisenberg model at the critical
point between the gapless spin fluid phase and the dimerized phase 
\cite{haldanej1j2}. The spin-spin correlations of this wavefunction have been 
computed 
exactly\cite{gebhard1,gebhard2}, but we are not aware of an exact 
result for its dimer-dimer correlations. We present numerical results for both
correlations below.
For $N=4$, given the possibility that the projected free Fermi gas wavefunction
at half-filling could be a candidate for the ${\cal C}-{\cal D}$ 
transition, we study spin-spin and dimer-dimer correlation functions of
the wavefunction and compare to exact results from the field theory 
for this transition. We also examine the
correlation functions in the projected Fermi gas wavefunction for $N>4$ 
as such wavefunctions may describe multicritical
points in the phase diagram of generalized $SU(N>4)$ spin models.

\begin{figure}
\centerline{\includegraphics[width=5in]{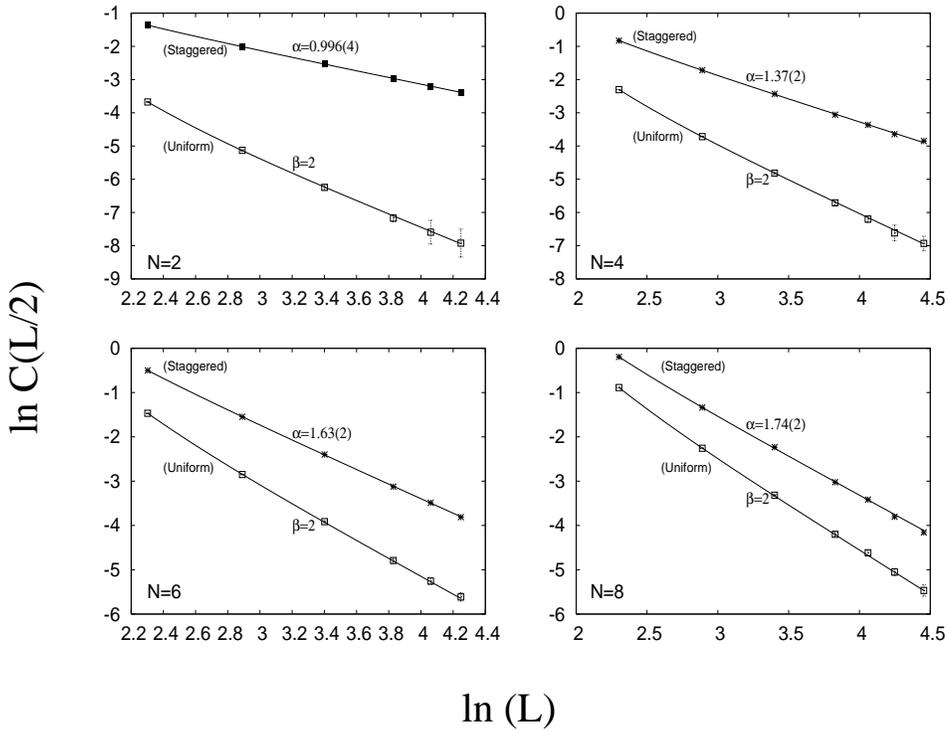}}
\caption{Logarithm of the staggered and uniform components of the spin-spin 
correlation function at distance $L/2$ plotted versus $\ln(L)$, for the 
projected free Fermi gas with $N$-flavors ($L$ denotes the number of sites
of the spin chain). The uniform spin correlations 
appear to decay as $1/r^2$ for all $N$. The staggered spin correlations 
are enhanced at smaller $N$, and decay with the indicated exponents 
(see text for details).}
\label{fig2}
\end{figure}

\ni\underline{Spin-spin correlations:}
At $N=\infty$, the long distance behavior of the spin correlation
function $C_{ss}(x) = \la S^\alpha_\beta(0) S^\beta_\alpha(x) 
\ra$ is given by mean field theory, $C_{ss}(x) \sim (-1/x^2 + (-1)^x/x^2)$. 
In the opposite limit, $N=2$, the spin correlations of the projected 
Fermi gas wavefunction have been calculated exactly by Gebhard and Vollhardt
\cite{gebhard1, gebhard2}
to be $C_{ss}(x) = (-1)^x \frac{3 Si(\pi x)}{2\pi x}$, where $Si(x)$ is the
sine integral function $Si(x)=\int_0^1 dy \sin (xy)/y$. The long distance 
decay of the correlator is $C_{ss}(x) \sim \frac{(-1)^x}{x} - 
\frac{2}{\pi^2 x^2}$. Thus the staggered spin correlations decay more
slowly for $N=2$ than they do in mean field theory, while the uniform
component continues to decay as $1/x^2$.

Incorporating gauge fluctuations perturbatively\cite{kim-lee,rantner} 
modifies the long distance spin correlations to:
\be
C_{ss}(x) \sim \frac{A (-1)^x}{x^{\alpha_s}}- \frac{B}{x^{\beta_s}}\ .
\ee
The ${\cal O}(1/N)$ result for the exponents are $\beta_s=2$
and $\alpha_s = 2-2/N$; rather surprisingly, the latter exponent agrees 
with the exact result.
Thus, gauge fluctuations enhance the staggered spin correlations 
over the mean field result, while the uniform component decays at the same 
rate $\sim 1/x^2$ as the mean field result.

In order to extract the exponent $\alpha_s(N)$ from the numerical 
calculations on the Gutzwiller projected wavefunction for
general $N$, we assume that the correlations are of the form
\be
C_{ss}(x) =  \frac{A_s (-1)^x}{x^{\alpha_s}} - \frac{B_s}{x^2}\ .
\ee
In order to carry out finite size scaling, we consider the function
\be
C(L/2) = (-1)^{L/2} C_{ss}(L/2) = \frac{A_s}{(L/2)^{\alpha_s}} 
- \frac{B_s (-1)^{L/2}}{(L/2)^2}\ .
\ee
From this equation we can obtain the staggered and uniform components of
the spin correlations via
\bea
C_{\rm stag}(L)&=&\frac{1}{2} \left[ C(L/2) + \frac{1}{2} \left( C(L/2+1)
+ C(L/2-1)\right) \right] \\
C_{\rm unif}(L)&=&\frac{1}{2} \left[ C(L/2) - \frac{1}{2} \left( C(L/2+1)
+ C(L/2-1)\right) \right]\ .
\label{stagunif}
\eea
To leading orders in $1/L$, these functions take the form
\bea
C^{\rm fit}_{\rm stag}(L)&\approx& \frac{A_s}{(L/2)^{\alpha_s}} +
\frac{A_s \alpha_s (1+\alpha_s)}{(L/2)^{2+\alpha_s}} + \frac{u_s}{(L/2)^4}\\
C^{\rm fit}_{\rm unif}(L)&\approx& \frac{B_s}{(L/2)^{2}} -
\frac{A_s \alpha_s (1+\alpha_s)}{L^2 (L/2)^{\alpha_s}} + \frac{v_s}
{(L/2)^4}
\eea
We also obtain the staggered and uniform components of the spin correlation
function data for various $L$ using equation \ref{stagunif}. We first fit the 
staggered correlations using the parameters $A_s,u_s,\alpha_s$. 
Next we use this value of $\alpha_s$ and the parameters $B_s,v_s$
to fit to the uniform component of the spin correlations.
The fits are shown in figure \ref{fig2} for both the uniform
and staggered components for cases $N=2$, $4$, $6$, and $8$. This leads to the following
estimates for various $N$:
$\alpha_s(2) = 0.996(4)$, $\alpha_s(4)=1.37(2)$, $\alpha_s(6)=1.63(2)$, and
$\alpha_s(8)=1.74(2)$. These values are in
remarkably good agreement with the exact value of the spin-spin correlation 
function exponent at the critical points of these chains, which for $N=4$ 
lies at the continuous ${\cal C}-{\cal D}$ transition.
We also find that the fitted value of the
amplitude ratio $A_s/B_s$ tends to unity as $N \to \infty$, 
consistent with mean field theory. At present, we are unable to
determine if the small differences between the exact
results and the wavefunction calculations of $\alpha_s(4)$ and $\alpha(6)$
are real or an artifact of working with chains of less than 100 sites.

\begin{figure}
\centerline{\includegraphics[width=5in]{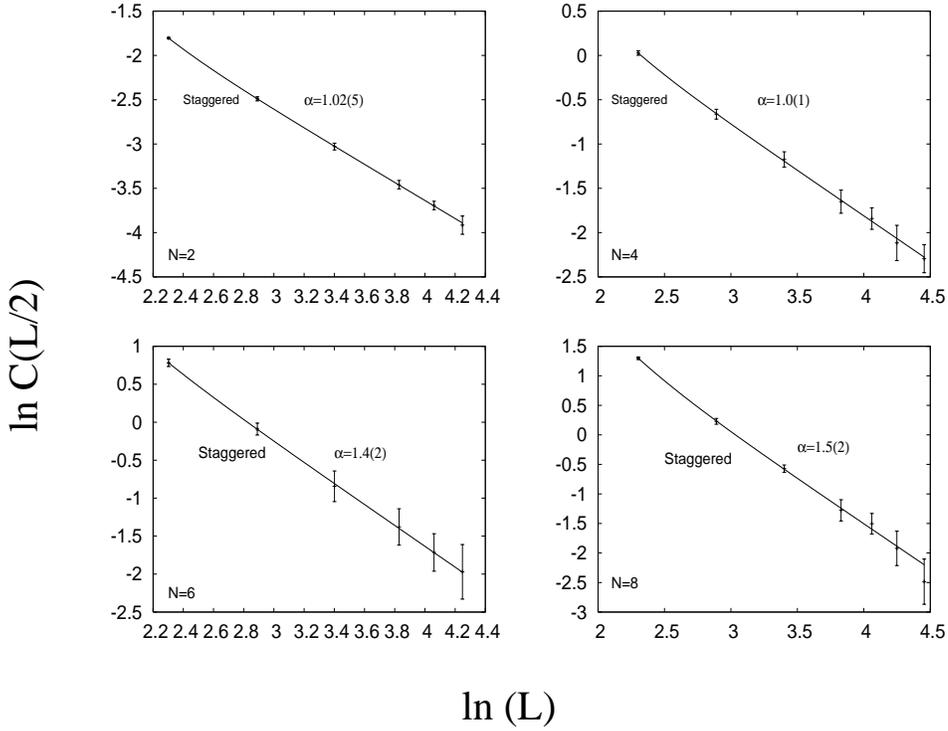}}
\caption{Logarithm of the alternating component of the dimer-dimer correlation 
function at distance $L/2$ plotted versus $\ln(L)$, for the 
projected free Fermi gas with $N$-flavors ($L$ denotes the number of
sites of the spin chain). The decay exponents are
indicated (see text). The decay of the uniform correlation is much 
faster and the corresponding exponents have not been obtained 
reliably.}
\label{fig3}
\end{figure}

\ni\underline{Dimer-dimer correlations:}
We have similarly evaluated the dimer-dimer correlations in the projected
$N$-flavor Fermi gas in 1D, namely $C_{dd}(x)=\la S^\alpha_\beta(0)
S^\beta_\alpha(1)S^\mu_\nu(x)S^\nu_\mu(x+1) \ra$. The finite size
scaling of this correlation function is analyzed in a manner
similar to that of the 
spin-spin correlation function, except for one significant difference. 
We fit to $C_{dd}(x) = A_d (-1)^x/x^{\alpha_d}$, dropping 
the uniform component that decays much more rapidly, so that we cannot 
extract its behavior reliably compared to the staggered component.  The finite 
size scaling plots of the
correlation function are shown in figure \ref{fig3}, along with estimates of 
$\alpha_d$
for various $N$. We find $\alpha_d(2)=1.02(5)$, $\alpha_d(4)=1.0(1)$,
$\alpha_d(6)=1.4(2)$, and $\alpha_d(8)=1.5(2)$. The projected Fermi 
gas wavefunction thus has strongly enhanced alternating
dimer correlations in addition
to enhanced staggered spin correlations. For $N=2$, it is in agreement
with the exact result $\alpha_d=2-2/N$. For $N>2$, the 
variational wavefunction exponents $\alpha_d(N) < \alpha_s(N)$ 
(most significantly for $N=4$) while
exact results suggest $\alpha_s(N)=\alpha_d(N)=2-2/N$. Thus, the
projected Fermi gas wavefunction does not quite capture this aspect of the
${\cal C}-{\cal D}$ critical point at $N=4$,
although it does capture the existence of strongly enhanced
staggered spin and alternating dimer correlations, decaying in
power-law fashion with anomalous exponents.
Having shown that the variational wavefunctions provide a reasonably good
description of $SU(N)$ spin models in 1D, we next turn to 2D
examples.

\section{Two Dimensional Square Lattice}
\label{2d}
Buoyed by the successful description of the $SU(N)$ spin chains with variational wavefunctions, 
we now apply the same methodology to the study of 2D $SU(N)$ spin models.   Much less 
is known reliably about the phase diagrams of such models at finite-$N$.   An interesting 
gapless spin liquid discovered some time ago in the large-$N$ limit of the self-conjugate 
$SU(N)$ Heisenberg model with biquadratic interactions to thwart dimerization is the 
$\pi$-flux state\cite{AM,MA}. At $N=\infty$, where projection to exactly $N/2$ fermions 
per site does nothing to the mean field state, the $\pi$-flux state corresponds to the 
ground state of fermions (at half-filling) hopping on the square lattice while 
sensing a (spontaneously generated) fictitious magnetic flux of $\pi$ per elementary 
plaquette\cite{AM}. It supports 
gapless linearly dispersing Dirac fermion excitations about two nodes in the
reduced Brillouin zone.  Spin-1 excitations, which are bilinears of the Dirac fermions,
are therefore gapless
at the wavevectors spanning the Dirac nodes: $(\pi, \pi)$, $(\pi, 0)$, $(0, \pi)$, and 
$(0, 0$).
For the case of the ordinary nearest-neighbor $SU(2)$ Heisenberg antiferromagnet 
on the square lattice, Gutzwiller projecting this mean field wavefunction leads to a
variational ground state with power law
decay of staggered spin correlations (as there is no long range magnetic order) 
and an energy $(1/2) \la S^\alpha_\beta(i) S^\beta_\alpha(i+\hat{\delta}) \ra 
\approx -0.319$ per bond. The true ground state of the Heisenberg model 
is known to be N\'eel ordered, with a spin moment of about 60\% of the classical value,
and $(1/2) \la S^\alpha_\beta(i) S^\beta_\alpha(i+\hat{\delta}) \ra \approx -0.3346$ 
per bond. Since the $\pi$-flux state is close to the true ground state, both energetically 
and in its display of (quasi) long-range antiferromagnetic correlations, we focus on the 
part of the phase diagram
of the $SU(N)$ Heisenberg model (with possible additional biquadratic interactions) that 
is close to that of the $\pi$-flux phase.  More precisely, 
we investigate instabilities of the $\pi$-flux state towards
various translational- and $SU(N)$-symmetry breaking orders.
This point of view was advocated in earlier studies of the $N=2$
case\cite{hsu,marston8}, and in more recent work by Ghaemi and Senthil \cite{ghaemi05}.

We begin with the $SU(N)$ Heisenberg model in the absence of
biquadratic interactions, and compare the resulting variational phase diagram with that from a recent quantum Monte Carlo (QMC) study of the same model\cite{assaad05}.  We then turn to the nature of the spin and dimer correlations of the $SU(4)$ projected
$\pi$-flux state and compare the correlation functions to results from recent analytical 
large-$N$ studies \cite{hermele05} and QMC 
calculations \cite{assaad05}.
Finally, we examine the variational phase diagram of the $SU(4)$ and $SU(6)$
models with a biquadratic 
interaction added to thwart instabilities.  For $N=6$, we find a 
(small) window of parameters where the projected $\pi$-flux state appears 
to be stable towards N\'eel, spin Peierls and broken-C 
ordering. Our work thus hints at the existence of a
{\em stable} $SU(6)$ gapless spin liquid phase in a simple two dimensional 
microscopic spin model.

\subsection{Phase Diagram of the $SU(N)$ Heisenberg model}
\label{2Dphase}

\begin{figure}
\centerline{\includegraphics[width=5.0in]{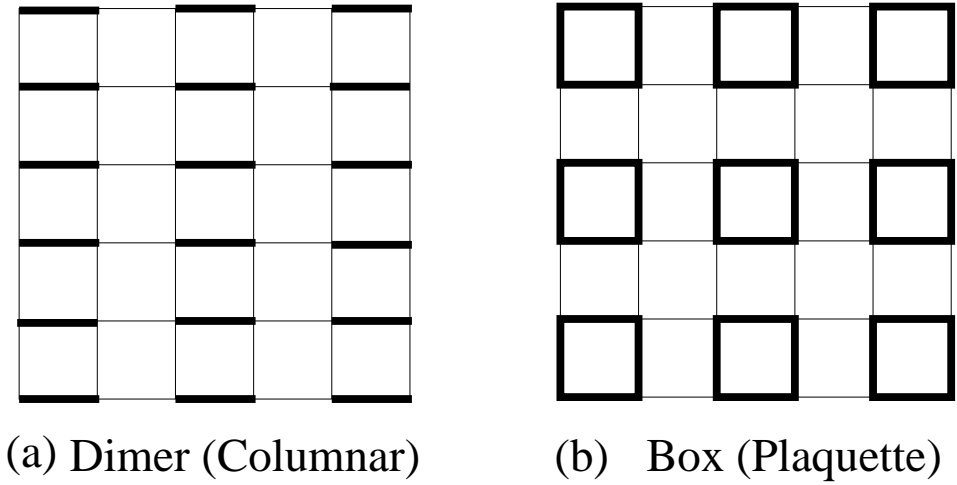}}
\caption{The two candidate spin Peierls ground states 
of the 2D $SU(N)$ spin models explored in this paper. The
thick lines indicate bonds with a larger singlet expectation 
value $|\la S^{\alpha}_\beta(i) S^{\beta}_\alpha(j)\ra|$.}
\label{fig4}
\end{figure}

The Hamiltonian of the 2D antiferromagnetic $SU(N)$ Heisenberg model is 
\be
H_{2D} = \sum_{\la i,j\ra} S^\alpha_\beta(i) S^\beta_\alpha(j),
\label{2dHeisenberg}
\ee
where $\la i,j \ra$ denotes nearest neighbor sites on the square lattice.
As discussed above, the projected $\pi$-flux wavefunction with $N/2$ 
fermions at each site is an attractive starting point to describe the 
ground state of the model.  The mean field ansatz for the $\pi$-flux phase is
\be
H_{\pi-{\rm flux}} = \sum_{\la i j\ra} t_{ij} \left( e^{i a_{ij}} 
f^{\dagger \sigma}_i f^{\vphantom\dagger}_{j\sigma} + {\rm h.c.} \right)
\ee
with $t_{ij}=t$ and a gauge choice of
$a_{i, i+\hat{x}} =  \frac{\pi}{4} (-1)^{x_i+y_i}$ and
$a_{i, i+\hat{y}} =  -\frac{\pi}{4}(-1)^{x_i+y_i}$.
Here we examine the
instability of the variational state obtained by Gutzwiller
projecting the ground state
of this mean field Hamiltonian towards N\'eel and spin Peierls ordering. (We
have also checked that there are no instabilities to time-reversal symmetry broken states or states with broken charge-conjugation symmetry.  These wavefunctions have higher energy, so the ground states exhibit only N\'eel or spin Peierls order.)

To account for the possibility of N\'eel ordering we modify the mean-field Hamiltonian
with the addition of a $SU(N)$ symmetry breaking perturbation that favors
two-sublattice ordering, with any chosen set of $N/2$ flavors favored on one sublattice, 
and the remaining $N/2$ flavors on the other sublattice:
\be
H_{neel}=H_{\pi-{\rm flux}} - h_{\cal N} \sum_{i,\sigma \leq N/2}
(-1)^{x_i+y_i} f^\dagger_{i\sigma} f_{i\sigma}
+ h_{\cal N} \sum_{i,\sigma > N/2}
(-1)^{x_i+y_i} f^\dagger_{i\sigma} f_{i\sigma}\ .
\ee
In order to study spin Peierls ordering, we focus on two different 
types of broken symmetry states, ``dimer order'' (more precisely, columnar 
dimer order) and ``box order'' (also called plaquette order). In the
dimer state, spins prefer to form singlets on nearest neighbor bonds,
and the bonds organize as shown in figure \ref{fig4}(a). 
Three other equivalent, but distinct, states are
obtained by $x-$translations and $\pi/2$ rotations of the displayed
pattern. 
The mean field ansatz for the preprojected wavefunction of the dimer state
is obtained by modulating $t_{ij}$ such that $t_{i,i+\hat{x}}=1 + \delta_D 
(-1)^{x_i}$ and $t_{i,i+\hat{y}}=1$. 
The box state has a different broken symmetry; the strength of
singlet bonds is shown in figure \ref{fig4}(b). Three other equivalent
box states are obtained by $x-$ and $y-$ translations of the displayed
pattern. For the box state, we modulate
$t_{ij}$ such that $t_{i,i+\hat{x}}=1 + \delta_B (-1)^{x_i}$ and 
$t_{i,i+\hat{y}}=1 + \delta_B (-1)^{y_i}$. 

\begin{figure}
\centerline{\includegraphics[width=5.0in]{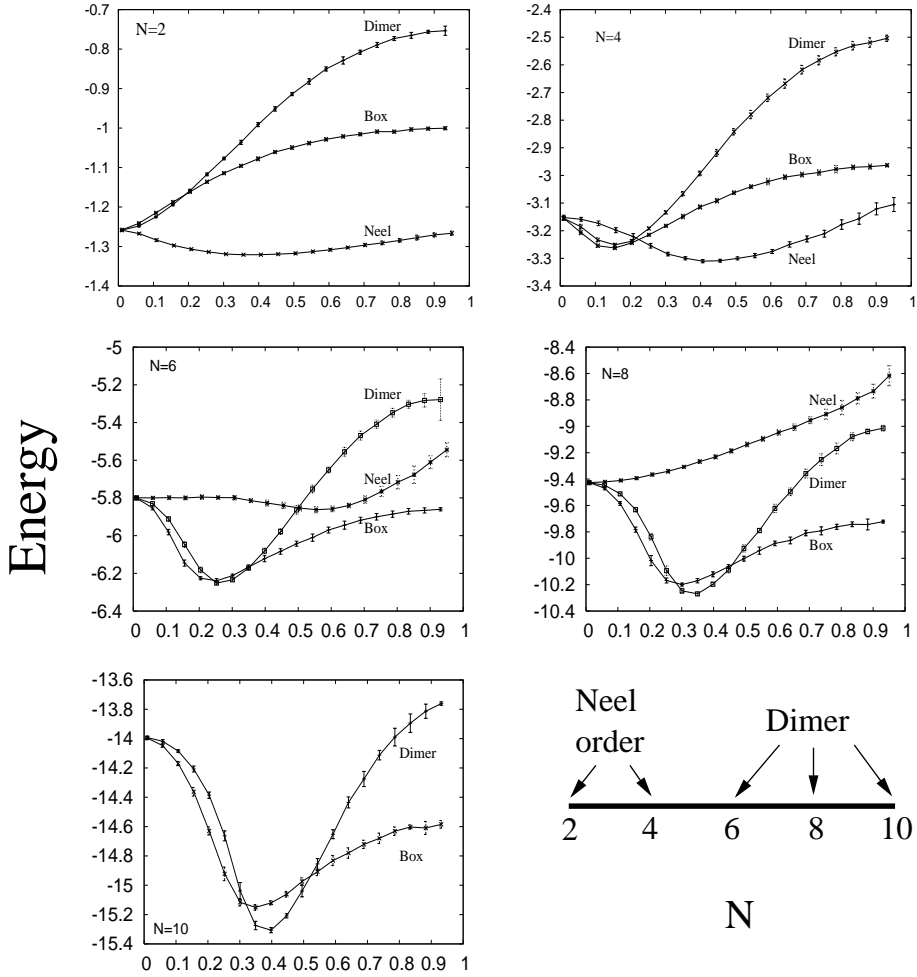}}
\caption{
Energy minimization plots for even $N=2$ to $10$.  The x-axis is the
variational parameter appropriate to the broken symmetry being
studied.  The y-axis is the (dimensionless) energy ($\la H_{2D} \ra$ for
the model of equation \ref{2dHeisenberg}). We conclude that the
$SU(N)$ Heisenberg model exhibits N\'eel order for $N=2$ and $4$, and
spin Peierls ordering of the columnar dimer type for $N > 4$.}
\label{fig5}
\end{figure}

The mean field box and dimer states have a single particle gap to fermionic 
excitations and 
thus also a spin gap. These spin Peierls ordered states as well as the 
N\'eel state are invariant
under a particle-hole transformation (followed by a global $SU(4)$ spin rotation
in the case of the N\'eel state), and thus are at half-filling.  We project these
mean field ansatz to obtain variational spin wavefunctions for the
Heisenberg model. The phases of the $SU(N)$ 
Heisenberg model are then obtained by looking for the state with the 
lowest variational energy. As summarized in figure \ref{fig5}, we find that 
the N\'eel ordered state has the
lowest energy for $N=2$ and $4$, while the dimer state (i.e., columnar dimer)
state has the lowest energy for $N > 4$. For $N=2$, this result is
in agreement with other numerical work \cite{trivedi1,trivedi2,
assaad05}. The presence of spin Peierls order for large values of $N$ 
is in agreement with
$1/N$ calculations \cite{AM}.  The same pattern of (columnar) dimer order 
was also predicted for various representations of $SU(N)$ antiferromagnets in 
large-N calculations by 
Read and Sachdev\cite{sachdev1,sachdev2}; these predictions have received 
some numerical support\cite{harada03}, with N\'eel order reported 
for $N \leq 4$ and dimer order for $N > 4$, identical to the phase diagram 
in figure \ref{fig5}.  

\subsection{Correlation Functions of the Projected $SU(4)$ $\pi$-Flux State}
\begin{figure}
\centerline{\includegraphics[width=5.0in]{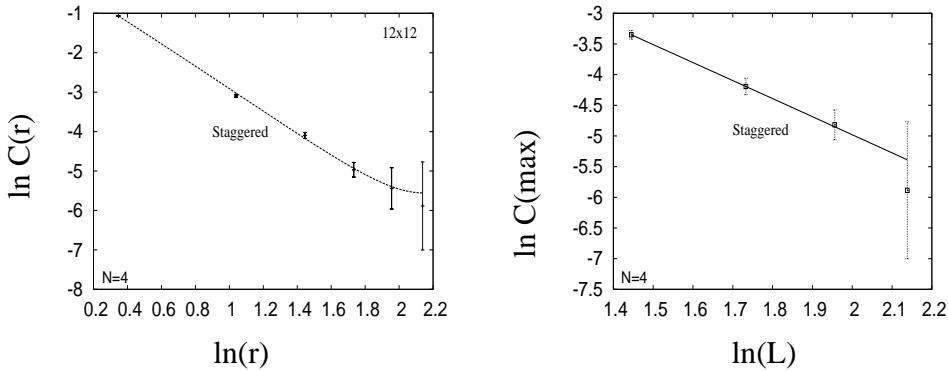}}
\caption{Spin-spin correlation function of the 2D projected $\pi$-flux
wavefunction for $N=4$ and a system with $L^2$ sites. The label
$C({\rm max})$ denotes the correlation function evaluated for points
with the maximal separation, $L/\sqrt{2}$, on the $L\times L$ square 
lattice.}
\label{fig6}
\end{figure}

To make contact with
a recent QMC study of the $SU(4)$ Heisenberg model \cite{assaad05},
we turn now to the spin-spin and dimer-dimer correlations of the projected
$\pi$-flux wavefunction. The analysis of these
correlations is done in a manner similar to that in 1D, except
that we focus only on the strong nonzero-wavevector component (near 
$(\pi, \pi)$ for the spin order and near $(\pi, 0)$ for the dimer correlations)
and ignore the uniform
components. The uniform component of the spin-spin correlation in 2D is
expected to decay quickly, as $\sim 1/r^4$, and is therefore numerically harder to evaluate.

The results for the finite size scaling of the staggered spin-spin
correlation function are shown in figure \ref{fig6}, together with the correlation
function results for the $12\times 12$ system. We find that the staggered
spin-spin correlation function decays as $1/r^\alpha$ with $\alpha_s(4)=
3.0(4)$. This value is in excellent agreement with large-$N$ calculations
\cite{hermele05,rantner} that find $\alpha_s(N) = 4 - 128/(3 \pi^2 N)$.
However, the exponent $\alpha_s(4)$ is much larger than that found in a QMC 
simulation by Assaad \cite{assaad05}.  Although the QMC calculation suggested
a spin liquid ground state of the $\pi$-flux type for the $SU(4)$ Heisenberg
model, Assaad found $\alpha_s \approx 1.12$. The origin of the large
difference needs further exploration, and we speculate on a possible reason 
for the discrepancy in the final section.

A numerical analysis of the dimer-dimer correlation function at
${\bf Q}=(\pi,0)$ for the
$N=4$ projected $\pi$-flux wavefunction yields
$\alpha_d(4)=2.1(8)$; the larger error on the dimer-dimer correlation 
function exponent stems from having fewer Monte Carlo samplings of the
wavefunction since the dimer-dimer correlation function takes more time
to evaluate.  We conclude
that Gutzwiller projection strongly enhances both the spin-spin and 
the dimer-dimer correlations relative to the mean field result. In this
sense, the projected $\pi$-flux state is indeed the ``mother of many 
competing orders'' \cite{hermele05}!
However, we have not confirmed yet that the 
projected $\pi$-flux wavefunction is an algebraic spin liquid
phase with enlarged $SU(2N)$ symmetry leading to $\alpha_d=\alpha_s$ 
\cite{hermele05}. We are currently carrying out further numerical 
calculations to reduce the error bars on the exponents and to better
test this prediction quantitatively.

\subsection{Adding Biquadratic Interactions for $N=4$ and $N=6$}
The inclusion of the biquadratic interaction lead to a rich phase diagram 
in the case of the $SU(4)$ spin chain.  Motivated by this physics, we 
pursue here a variational study of the 2D square lattice model with 
Heisenberg bilinear and biquadratic interactions for $SU(4)$ and $SU(6)$.
As in 1D, the 2D model is defined by the nearest-neighbor Hamiltonian:
\be
H_4 = \cos\theta \sum_{\la i,j\ra} S^\alpha_\beta(i) S^\beta_\alpha(j)
+ \frac{\sin\theta}{4} \sum_{\la i,j\ra} [S^\alpha_\beta(i) 
S^\beta_\alpha(j)]^2,
\label{su4d2}
\ee
where $\la i,j\ra$ refer to nearest neighbor sites on the 2D square
lattice.  Exact diagonalization and the density matrix
renormalization group are inadequate tools to study the phase diagram
of the two dimensional model.  For spin Hamiltonians without a sign problem,
QMC has proved to be the most reliable numerical tool in 2D, but the case of a 
positive biquadratic spin interaction cannot be studied reliably because 
it is a frustrating interaction that introduces the sign problem.  Given the success of the 
variational approach in 1D we have reason to hope that the method may
also provide a good guide to two dimensions,
although we explore only
a limited class of variational states.  We consider the fully
polarized ferromagnet ${\cal FM}$, the (columnar) dimer state ${\cal D}$,
the N\'eel antiferromagnet ${\cal N}$, the broken charge-conjugation symmetry state ${\cal C}$, and the projected $\pi$-flux state $\Pi$.

\noindent \underline{${\cal FM}$, Ferromagnet:}
The ground state of a fully polarized ferromagnet in 2D breaks 
global $SU(N)$ spin symmetry but no lattice symmetries. An exact
ground state wavefunction may be constructed by placing any two of the four 
fermion flavors on each lattice site, each site having the same two flavors. 
All other ground states can be obtained by global $SU(N)$ rotations of 
this state. Exactly as in 1D, because the Pauli exclusion principle 
prevents any fermion hopping in the ${\cal FM}$ state, it is straightforward 
to show that $\la S^\alpha_\beta(i) S^\beta_\alpha(i+1) \ra = N/4$ and
$\la [S^\alpha_\beta(i) S^\beta_\alpha(i+1)]^2 \ra = N^2/16$,
so that the exact ground state energy per site is 
\be
e^{\rm exact}_{\cal FM}(N) = 
\frac{N}{2} \cos\theta + \frac{N^2}{32} \sin\theta.
\ee

\noindent \underline{${\cal D}$, Dimer:}
The columnar dimer state appeared in the phase diagram
of the pure bilinear Heisenberg model and hence was discussed above in subsection \ref{2Dphase}. The only difference here is
that we minimize the energy with respect to $\delta_D$ at each value of
$\theta$. The energy of the perfectly dimerized state for $N=4,6$ is
\bea
e^{\rm var}_{\cal D}(\delta_D=1,N=4)  
&=& -\frac{5}{2} \cos\theta + \frac{15}{4} \sin\theta \\
e^{\rm var}_{\cal D}(\delta_D=1,N=6)  
&=& -\frac{21}{4} \cos\theta + \frac{1197}{80} \sin\theta.
\eea

\noindent \underline{${\cal N}$, N\'eel:}
The N\'eel state with long range antiferromagnetic order was also discussed above.
Here we minimize the energy with respect to the staggered 
magnetic field $h_{\cal N}$ at each $\theta$. We emphasize that the
classical antiferromagnetic state (obtained in the limit $h_{\cal N}\to \infty$)
has a much higher energy than the optimal antiferromagnetic ground 
state at finite $h_{\cal N}$ in the relevant region of the
phase diagram.

\noindent \underline{${\cal C}$, Charge-conjugation symmetry broken:} The
${\cal C}$ phase is obtained by projecting the ground state of the
mean field Hamiltonian
\be
H^0_{{\cal C}}=
H_{\pi-{\rm flux}} 
- t^* \sum_{i,\alpha=1\ldots 4} (-1)^{x_i+y_i}
\left[f^{\dagger \alpha}_i f_{i+2\hat{x} \alpha}  + 
f^{\dagger \alpha}_i f_{i+2\hat{y} \alpha} + {\rm h. c.} 
\right]\ ,
\ee
the 2D generalization of equation \ref{c-break}.  
The intra-sublattice hopping $t^*$ gaps out the Dirac nodes of fermionic 
excitations in the $\pi$-flux state and thus leads to a spin gap in the 
mean field spectrum.

\noindent \underline{$\Pi$, The $\pi$-flux state:}
This gapless spin liquid state was also introduced earlier. This
state does not have any variational parameters and is thus the most
constrained state of the wavefunctions that we study. We leave
for future work possible variational modifications of the state that preserve all 
lattice and spin symmetries. Since the $\pi$-flux
state is a gapless state, it is important to study it under conditions
such as particular system sizes that permit gapless nodes to appear at the 
mean field level.  We find that the phase is artificially 
stabilized on lattices that do not permit the nodal wavevectors.

\begin{figure}
\centerline{\includegraphics[width=6in]{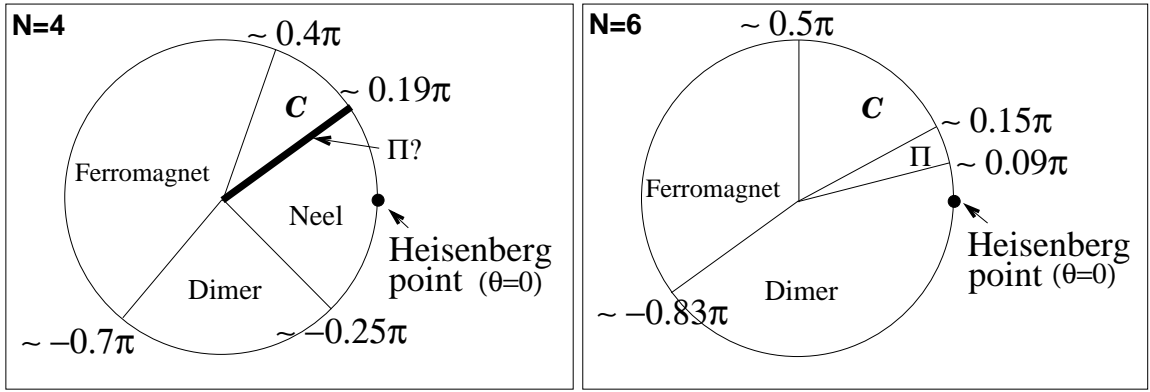}}
\caption{Phase diagram of the $SU(4)$ and $SU(6)$ spin models
of equation \ref{su4d2}, with Heisenberg bilinear and biquadratic interactions, 
on the 2D square lattice. Phases appearing here are the N\'eel
phase, the ferromagnet, the columnar dimer phase, the broken charge-conjugation
symmetry phase (${\cal C}$), and the $\pi$-flux spin-liquid phase ($\Pi$).
The thick line labeled ``$\Pi$?'' in the $N=4$ phase diagram
indicates that the projected $\pi$-flux state could either be stable
over a thin sliver region ($0.18 < \theta/\pi < 0.20$) or instead there 
may be a direct transition at $\theta/\pi \approx 0.19$
between the N\'eel and ${\cal C}$ phases. See text for details.}
\label{fig7}
\end{figure}

\subsection{Phase diagram of the $SU(4)$ spin model}
We obtain the variational phase diagram shown in the left panel of figure \ref{fig7} 
from an evaluation of the energy of the various $SU(4)$ states. 
The ground state appears to generically exhibit broken
symmetry. The ${\cal D}-{\cal FM}$, ${\cal D}-{\cal N}$
and the ${\cal C}-{\cal FM}$ transitions appear strongly first order due to
level crossings. As in 1D, the ground state is nearly completely
dimerized at the ${\cal D}-{\cal FM}$ transition. The 
location of the ${\cal D}-{\cal FM}$ 
transition is therefore simply and reliably estimated by studying a dimer 
product wavefunction as the variational state for ${\cal D}$.
Setting $e^{\rm exact}_{\cal FM} = e^{\rm var}_{\cal D}(\delta_D=1)$
leads to $\theta^{\rm var}_{{\cal D}-{\cal FM}} \approx \tan^{-1}(18/13) \approx
-0.7\pi$.

Due to limits on numerical accuracy and on system sizes (up to $10 \times 10$ 
for variational optimization), we are unable to determine 
whether there is a direct ${\cal N}-{\cal C}$ transition at $\theta/\pi \approx 0.19$ or a thin
sliver of the $\pi$-flux phase that intervenes between these two phases for $0.18 < \theta/\pi < 0.20$.
Since the N\'eel and ${\cal C}$ states are both
deformations of the $\pi$-flux state, it is possible for a direct continuous transition to occur
between them, with the projected
$\pi$-flux state being a possible candidate for the critical point. 
However the addition of 
variational parameter(s) to improve short distance correlations of the
projected $\pi$-flux state could in fact stabilize this state.
We are examining this issue more carefully, and comment further on this 
point in the final section.  

\subsection{Phase diagram of the $SU(6)$ spin model}
The variational phase diagram is shown in the right panel of
figure \ref{fig7}.  Strikingly the N\'eel phase is completely replaced 
by the dimerized phase at $N=6$, and the biquadratic
interaction appears to stabilize the $\pi$-flux state over
a small window of $\theta$. Of course within the variational approach 
one cannot rule out the possibility that $\Pi$ may be unstable to some 
other more complicated or exotic broken symmetries that have not been considered.

The ${\cal D}-{\cal FM}$, ${\cal D}-{\cal N}$
and the ${\cal C}-{\cal FM}$ transitions appear strongly first order again due to 
level crossings. Since the ground state is nearly completely
dimerized at the ${\cal D}-{\cal FM}$ transition the 
location of this
transition is reliably estimated by studying a dimer 
product wavefunction as the variational state for ${\cal D}$.
Setting $e^{\rm exact}_{\cal FM} = e^{\rm var}_{\cal D}(\delta_D=1)$
leads to $\theta^{\rm var}_{{\cal D}-{\cal FM}} \approx \tan^{-1}(660/1107) 
\approx -0.83\pi$.

\section{Summary and Discussion}
\label{discuss}

We have used Gutzwiller projected variational wavefunctions
to deduce phase diagrams of $SU(N)$ antiferromagnets with Heisenberg bilinear
and biquadratic interactions in one and two spatial dimensions. In 
one dimension,
the $SU(4)$ variational phase diagram is in very good agreement with exact
results. The spin and dimer correlations of the projected Fermi gas
wavefunction with $N$ fermion flavors are also in reasonably
good agreement with $1/N$ calculations and exact results. 
Based on these results, the projected free fermion state 
with $N$ fermion flavors appears to 
provides a good approximation of the critical points of $SU(N)$ spin chains,
and in particular it is a good description of the critical point between 
dimerized and broken 
charge-conjugation symmetry phases in the $SU(4)$ model.

On the two dimensional square lattice the pure bilinear Heisenberg model  
exhibits N\'eel order for $N=2$ and $4$ and columnar dimer order 
for $N > 4$.  Biquadratic interactions of positive sign
appear to destabilize the N\'eel state, as the N\'eel order diminishes
and gives way to a broken charge-conjugation symmetry phase via
either a small sliver of the $\pi$-flux spin liquid or by a continuous 
transition that is well described by the projected $\pi$-flux state. 

The spin and dimer
correlations of the projected $SU(4)$ $\pi$-flux state are in reasonable
agreement with analytical $1/N$ calculations. 
However the spin-spin correlations
are quite different from those reported in QMC calculations by
Assaad \cite{assaad05}. 
While that study finds a spin liquid ground
state for the $SU(4)$ Heisenberg model, apparently of the $\pi$-flux type, 
the spin correlations decay {\em much}
more slowly than those predicted on the basis of $1/N$ calculations or 
our variational calculation. Based on the variational study of model equation \ref{su4d2}
with biquadratic
interactions, we find that there could either be a thin sliver of
the flux phase or a direct continuous N\'eel-${\cal C}$ transition 
at $\theta/\pi \approx 0.19$.  In the exact phase diagram this transition point, or the sliver of the
flux phase, might occur even closer to the pure bilinear Heisenberg point. If this in fact is the case,
a continuous N\'eel-${\cal C}$ transition with a 
$\pi$-flux state at the transition point (or a direct N\'eel-$\Pi$ 
transition if a region of stable $\pi$-flux spin liquid exists) 
could strongly influence the ground state of the pure $SU(4)$
Heisenberg model as studied by QMC \cite{assaad05}. 
This hypothesis suggests that it may be numerically difficult to 
tell whether the correct ground state of the $SU(4)$ Heisenberg model is
a $\pi$-flux spin liquid or a N\'eel ground state with a much reduced 
staggered magnetization. It could also
account for the discrepancy in the spin correlations 
between the QMC on one hand and the $1/N$ and variational calculations
on the other. 
Further studies of the $SU(4)$ Heisenberg model with biquadratic 
interactions might shed light on this issue.  

The $SU(6)$ model does not appear to support a N\'eel phase at all. 
Instead biquadratic interactions open up a small window of $\pi$-flux
phase between the dimerized and broken charge-conjugation symmetry
phases. 
We checked for instabilities of the $\pi$-flux state towards 
N\'eel order (characterized by non-zero $\la S^\alpha_\beta(i) \ra$), 
dimer order (modulations in 
$\la Tr S(i) S(j) \ra$) and C-breaking (modulations in $\la Tr S(i) S(j) S(k) 
\ra$) and found it to be stable against all three.  
However, we cannot rule out instabilities towards other more
exotic broken symmetries characterized by more complicated order 
parameters. While the dimerized ground state at Heisenberg point also 
appears to be 
close to a spin liquid phase from our phase diagram, it is less likely to be 
influenced by proximity to such a critical point as the
dimerized phase has a spin gap rendering it more stable to critical
fluctuations than the N\'eel phase. This picture
is consistent with Assaad's QMC results,
with a dimerized ground state reported for the $SU(6)$ Heisenberg model.

Finally, an exact C-breaking ground state of
a 2D $SU(8)$ spin model is known at a special point in parameter space
\cite{arovas} and it
could be used as an additional test of the variational approach, which
we have shown to be quite successful in describing a wide class of
$SU(N)$ antiferromagnets in one and two dimensions.

\ack{We thank F. Assaad, P. Fendley, M. Hermele, P. Lee, and T. Senthil for 
useful discussions and correspondence.
This work was supported in part by the National Science Foundation under 
grant No. DMR-0213818 (JBM) and by the National Sciences and Engineering
Research Council of Canada (AP).  This work was initiated during the 
``Exotic Order and Criticality in Quantum Matter'' program at 
the Kavli Institute for Theoretical Physics supported in part by the NSF 
under grant No. PHY99-074. AP also thanks the Aspen Center for Physics
Summer Program on ``Gauge Theories in Condensed Matter Physics'' where
part of this work was completed.}

\end{document}